\documentclass[aip,graphicx]{revtex4-2}
\usepackage[hidelinks]{hyperref}
\usepackage{graphicx}% Include figure files
\usepackage{xcolor}
\usepackage{ulem}

\draft % marks overfull lines with a black rule on the right

\begin{document}

% Use the \preprint command to place your local institutional report number 
% on the title page in preprint mode.
% Multiple \preprint commands are allowed.
%\preprint{}

\title{Anisotropic frictional model for an object sliding in a granular media\footnote{Please cite this article as DOI:~\href{https://doi.org/10.1063/5.0188244}{https://doi.org/10.1063/5.0188244}}} %Title of paper

\email[Corresponding author: ]{rene.zuniga@usach.cl}

\author{Ren{\'e}  Zu{\~n}iga$^{\ast}$}
\affiliation{Department of Physics, Universidad de Santiago de Chile, Avenida Ecuador 3493,Estaci\'on Central, 9170124, Santiago, Chile.} 
\affiliation{Center for Soft Matter Research, SMAT-C, Avenida Bernardo O’Higgins 3363,Estaci\'on Central, 9170124, Santiago, Chile.}

\author{Carlos Vasconcellos}
\affiliation{Department of Physics, Universidad de Santiago de Chile, Avenida Ecuador 3493,Estaci\'on Central, 9170124, Santiago, Chile.} 
\affiliation{Center for Soft Matter Research, SMAT-C, Avenida Bernardo O’Higgins 3363,Estaci\'on Central, 9170124, Santiago, Chile.}

\author{Baptiste Darbois Texier}
\affiliation{Universite Paris-Saclay, CNRS,Laboratoire FAST, 91405, Orsay, France.}

\author{Francisco Melo}
\affiliation{Department of Physics, Universidad de Santiago de Chile, Avenida Ecuador 3493,Estaci\'on Central, 9170124, Santiago, Chile.} 
\affiliation{Center for Soft Matter Research, SMAT-C, Avenida Bernardo O’Higgins 3363,Estaci\'on Central, 9170124, Santiago, Chile.}

% Collaboration name, if desired (requires use of superscriptaddress option in \documentclass). 
% \noaffiliation is required (may also be used with the \author command).
%\collaboration{}
%\noaffiliation

\date{\today}

\begin{abstract}
Several locomotion strategies are based on the anisotropic nature of the forces experienced by the moving body with its environment. We report experiments on the anisotropy of the frictional force experienced by a cylinder moving in a granular medium as a function of the orientation $\alpha$ between the cylinder and its velocity. The component of the force in the direction parallel to the velocity of the cylinder is always higher than the perpendicular component and the force is therefore anisotropic. While the parallel component increases continuously with the cylinder angle $\alpha$, we observe that the perpendicular component reaches a maximum value for an orientation of $\alpha_c \simeq 35^\circ$. In order to rationalise these observations, we have developed a theoretical model which assumes that the mechanical energy required to move the cylinder is dissipated by friction and establishes a relationship between the parallel and perpendicular force components that is consistent with experiments.
\end{abstract}

% insert suggested PACS numbers in braces on next line

\maketitle %\maketitle must follow title, authors, abstract and \pacs

% Body of paper goes here. Use proper sectioning commands. 
% References should be done using the \cite, \ref, and \label commands
%\section{}
%\label{}
%\subsection{}
%\subsubsection{}

%%%%%%%%%%%%%%%%%%%%%%%%%%%%%%%%%%%%%%%%%%%%%%%%%%%%%%%%%%%%%%%%%%%%%%%%%%%%%%%%%%%%%%%%%%%%%%%%%%%%%%%%%%%%%%%%%%%%%%%%%%%%
\section{Introduction}\label{Intro}

Locomotion is an essential aspect of life for many species, and living beings have adopted different strategies depending on their morphology and environment. Micro-organisms present in fluids use locomotion strategies based on the movement of a slender structure for which the viscous forces are known to be higher in the transverse direction of the structure than in the tangential direction \cite{lauga2009hydrodynamics}. In these strategies, the anisotropy of viscous forces is coupled with non-reciprocal motion such as the rotatory motion of cilia as helical propulsion \cite{Brennen1977,Polin2009,DeLima2011,Rode2021}, wave propagation (e.g. C. Elegans) \cite{Stephens2008,Xu2018,Zhang2019}, or the propagation of metachronal waves on an array of cilia (e.g. Paramecium) \cite{Guirao2007,Tsumori2016,Dong2020,Meng2021}. On a larger scale and on solid ground, several locomotion strategies of animals are based on the anisotropic nature of friction with the substrate. This is the case of snakes that propel on flat surfaces by slithering undulations \cite{Hu2009}. Hu \textit{et al.} have shown that the presence of scales under the snake's body generates a higher coefficient of friction for normal movement than for tangential movement, which is responsible for the snake's propulsion. The anisotropic friction is also at the origin of the propulsion into granular media such as sand. Maladen \textit{et al.} studied the propulsion of sandfish lizards below the sand surface and showed that it is due to the undulatory movement of the lizard and to the fact that the drag forces experienced in a granular medium are higher for displacement in the direction transverse to the body than for displacement along the body \cite{Maladen2009}. Inspired by this mechanism, Maladen \textit{et al.} designed a robot (the sand-swimming robot) capable of moving below the surface of the sand \cite{Maladen2011}. Using the anisotropic nature of forces in a granular medium, Darbois Texier \textit{et al.} have developed a robot with a rotating helix that can swim through this material \cite{Darbois-Texier2017}. Beyond the problem of locomotion, the forces experienced by an object moving in grains have been extensively studied in different configurations. This is the case, for example, of the granular drag of a cylindrical intruder moving in the axial and lateral direction \cite{aghda2023drag}, the lift experienced by a cylinder moving horizontally \cite{guillard2014lift}, the drag and lift experienced by an object immersed in a gravity-driven flow down an inclined plane \cite{yennemadi2023drag}. These force measurements in granular materials have led to the adaptation of the Resistive Force Theory (RFT), developed to rationalize the forces experienced by a slender object into viscous fluids \cite{Gray1955}, to the case of granular materials \cite{Zhang2014}. It is well known that the description of granular flows around objects is of great complexity \cite{Cui2022,Mandal2019}, due to the absence of constitutive relations that allow to close the force balance and mass conservation equations.  To reduce such complexity, we propose here a simplified description for forces on cylindrical objects based on a maximum energy release rate condition. In the same vein, the granular RFT has shown a good level of effectiveness to predict granular forces in different configurations \cite{Hatton2013,askari2016intrusion,Peng2016,Agarwal2021,Liu2021,Huang2022,Yuan2023}. However, the mathematical description of the granular forces experienced by an intruder remains empirical rather than based on physical principles, particularly when it comes to describing their anisotropic nature as a function of the object orientation, which is important for locomotion problems.\\

In this article, we investigate the physical foundation of anisotropic friction forces in a granular medium. We realize experiments of a cylinder dragged into a granular material with different orientations and depths. We first detail the setup and the experimental protocol followed in this study. Section III presents the results of anisotropic friction experienced by the cylinder. Finally, Section IV details a model based on energetic considerations to rationalise our observations.\\ 

%%%%%%%%%%%%%%%%%%%%%%%%%%%%%%%%%%%%%%%%%%%%%%%%%%%%%%%%%%%%%%%%%%%%%%%%%%%%%%%%%%%%%%%%%%%%%%%%%%%%%%%%%%%%%%%%%%%%%%%%%%%%
\section{Experimental setup}\label{sec:setup}

%%%%%%%%%%%%%%%%%%%%%%%%%%%%%%%%%%%%%%%%%%%%%% FIGURE 1 %%%%%%%%%%%%%%%%%%%%%%%%%%%%%%%%%%%%%%%%%%%%%%%%%%%%%%%%%%%%%%%%%%%%
\begin{figure}[t] 
\centering
\includegraphics[width=0.5\columnwidth]{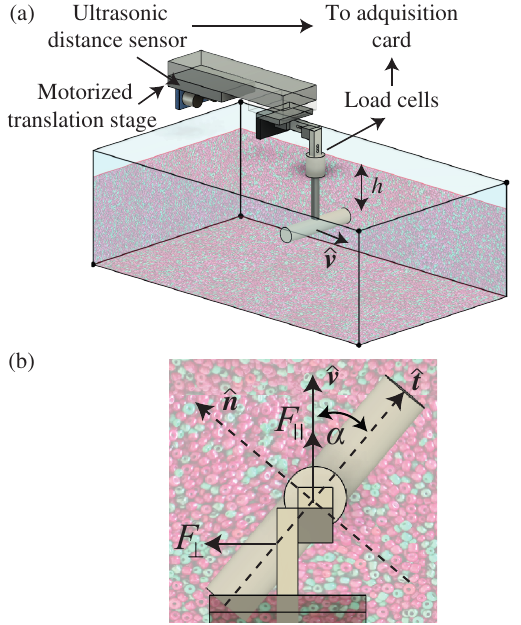}
\caption{(a) The experimental setup developed to measured the parallel and perpendicular forces experienced by a cylinder moving into a granular medium. (b) Top view of the dragged cylinder with the experimental parameters and measured quantities.}
\label{fig:setup}
\end{figure}
%%%%%%%%%%%%%%%%%%%%%%%%%%%%%%%%%%%%%%%%%%%%%%%%%%%%%%%%%%%%%%%%%%%%%%%%%%%%%%%%%%%%%%%%%%%%%%%%%%%%%%%%%%%%%%%%%%%%%%%%%%%%

The experimental setup consists of a cylinder dragged through a granular medium and is illustrated in Fig. \ref{fig:setup}(a). The dragged cylinder has a length $L=80$ mm, a diameter $D = 15$ mm, and is inclined by an angle $\alpha$ relative to the direction of movement. It is connected to two load cells (Phidgets model 3132) placed above the grains by a thin steel rod of 8 mm in diameter. The surface area of the steel rod is chosen to be small compared with the surface area of the cylinder and to generate negligible drag. One of the sensors is oriented to measure the force component parallel to the velocity $F_{\parallel}$, and the other to measure the force component perpendicular to the velocity $F_{\perp}$, as shown in Fig. \ref{fig:setup}(b). The orientation angle $\alpha$ of the cylinder can be changed by simply rotating the thin connecting rod. The cylinder is moved horizontally using a controlled translation stage (Thorlabs model MTS50/M-Z8), which allows constant-speed movement to be imposed over a speed range from $v=0.5$ mm/s to $v=2.4$ mm/s and a maximum displacement of $x_{\rm{max}}=50$ mm. The displacement is measured using an ultrasonic sensor (model Hc-sr04) operating at a frequency of 100,000 Hz and with a spatial resolution of 0.1 mm.

\noindent The granular medium filling the container is made up of glass beads with an average diameter of $d_g = 3$ mm and bulk density $\rho_g = 2200$ kg/m$^3$. The mean avalanche angle of the granular material is $31.3^\circ \pm 1.3 ^\circ$. Before each experiment, we prepare the granular medium to obtain the same initial packing fraction of $\phi=0.61 \pm 0.02$. The container is large enough (300 mm $\times$ 200 mm $\times$ 200 mm) to prevent edge influence and Janssen effects \cite{seguin2008influence,bertho2003dynamical}. The depth of the cylinder in the granular medium, $h$, is controlled by lifting the container using a lift table mechanism, where the depth $h=0$ mm corresponds to when the cylinder lightly touches the first layer of grains, the maximum depth used is $h=50$ mm. 
%%%%%%%%%%%%%%%%%%%%%%%%%%%%%% FIGURE 2 %%%%%%%%%%%%%%%%%%%%%%%%%%%%%%%%%%%%%%%%%%%%%%%%%%%%%%%%%%%%%%%%%%%%%%%%%%%%%%%%%%%%
\begin{figure*}[t] 
\centering
\includegraphics[width=1\columnwidth]{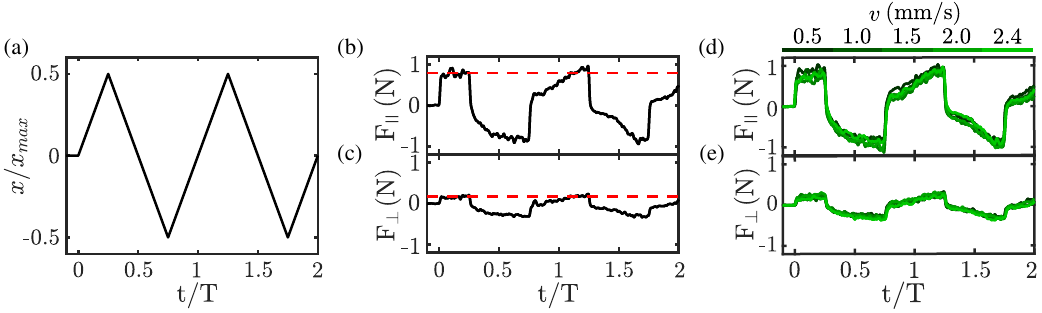}
\caption{A typical panel of signals is obtained for an experiment when the angle is $\alpha = 30^\circ$, the depth is $h=10$ mm, and the speed $v=1$ mm/s. From the signal obtained by the displacement sensor, (a) Displacement normalized by the maximum displacement of the cylinder through the granular medium $x_{max} =50$ mm as a function of time normalized by the effective period of a cycle $T\approx 100$ s. From the measured signals of the load cells in the direction parallel and perpendicular to the velocity, (b) Parallel force as a function of time, and (c) Perpendicular force as a function of time. The red dashed lines in (b) and (c) correspond to the average values of the force during the first displacement respectively. Experimental data shown were measured for constant angle and depth ($\alpha = 30^\circ$ and $h= 10$ mm) and different speeds (green lines), (d) Parallel force as a function of time, and (e) Perpendicular force as a function of time.}
\label{fig:signals}
\end{figure*}
%%%%%%%%%%%%%%%%%%%%%%%%%%%%%%%%%%%%%%%%%%%%%%%%%%%%%%%%%%%%%%%%%%%%%%%%%%%%%%%%%%%%%%%%%%%%%%%%%%%%%%%%%%%%%%%%%%%%%%%%%%%%
\noindent Once the depth and angle of the cylinder are fixed, the cylinder is positioned at half of the maximum displacement $x_{\rm{max}}$ and then moved back and forth in several cycles over a period of time $T$. From signals processing acquired by the ultrasonic distance sensor, we obtain the displacement of the cylinder as a function of time (Fig. \ref{fig:signals}(a)), which confirms that the movement takes place at constant speed. The parallel and perpendicular forces experienced by the cylinder during its motion are show in Fig. \ref{fig:signals}(b) and (c) respectively. When set in motion, the forces increases rapidly and reaches a plateau (red dashed line in Fig.\ref{fig:signals}(b-c)) while undergoing significant variations around this plateau. These variations result from the sliding of the material due to compression resulting from the passage of the cylinder, as noted previously \cite{Zhang2014,Darbois-Texier2021}. We observe that the parallel force is higher than the perpendicular one. After the first movement, the cylinder returns in its wake and the shape of the force curves changes, with the increase in force to reach the plateau taking longer over successive cycles. The evolution of the granular forces experienced by an objet moving in its own wake was previously reported by Guillard \textit{et al.} in the case of a rotating cylinder in a granular medium \cite{Guillard2013}, by Darbois Texier \textit{et al.} for a reciprocal motion of a rigid rod rotating in a granular medium \cite{Darbois-Texier2021}, by Tapia \textit{et al.} for shear bands in granular flows induced by the penetration of an intruder \cite{Tapia2013}, and Vivanco \textit{et al.} for stress reduction by friction mobilization \cite{Vivanco2016}. In addition, we found that for the range of velocities possible to impose with the stage, the force curves present insignificant changes (Fig. \ref{fig:signals}(d-e)) indicating that these experiments are in the quasi-static granular force regime \cite{DeBlasio2009,Geminard1999,VanWal2017,Khala2021}. This observation is consistent with the estimate of the Froude number which is ${\rm Fr}=v/\sqrt{g(h+D/2)}\approx 0.0040 \pm 0.0015$ i.e. less than 0.1, the limit of the quasi-static regime. In the following, we present results of experiments made at constant velocity $v=2$ mm/s and focus on the forces experienced during the first displacement. For each measurment, we have made five repetitions in order to estimate the dispersion of the data.

%%%%%%%%%%%%%%%%%%%%%%%%%%%%%%%%%%%%%%%%%%%%%%%%%%%%%%%%%%%%%%%%%%%%%%%%%%%%%%%%%%%%%%%%%%%%%%%%%%%%%%%%%%%%%%%%%%%%%%%%%%%%

\section{Results}\label{sec:results}
We describe the behavior of the parallel and perpendicular forces as a function of the cylinder depth and its orientation angle $\alpha$ (see Fig. \ref{fig:setup}(b)).

\subsection{Paralell force}\label{subsec:parallel_force}

\begin{figure}[t] 
\centering
\includegraphics[width=1\columnwidth]{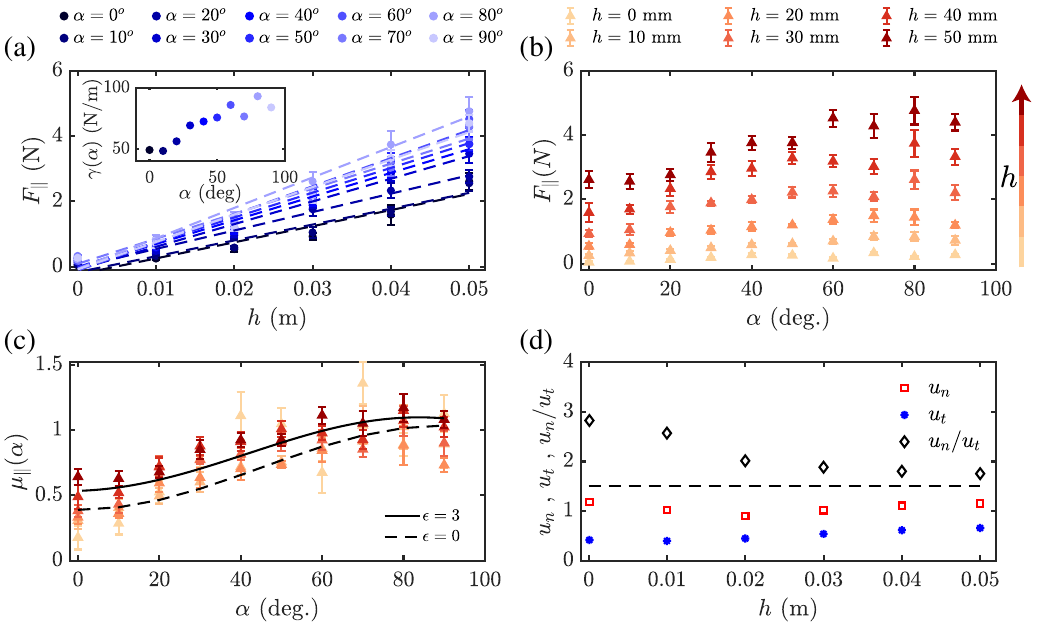}
\caption{(a) Parallel force as a function of depth for different angles $\alpha$, insert corresponds to the slopes of each curve as a function of the angle. (b) Parallel force as a function of the angle, the colours of the symbols code the depth. (c) Parallel friction coefficient as a function of angle, segmented black line corresponds to the theoretical friction coefficient without corrections (Eq. (\ref{eq7}) with $\epsilon = 0$) and solid black line for the theoretical friction coefficient with corrections (Eq. (\ref{eq7}) with $\epsilon = 3$). (d) coefficients $u_n$ (red squares), $u_t$ (blue asterisks) and the ratio $u_n/u_t$ (black diamonds) as a function of depth, segmented black line indicates the value 1.5 for the ratio $u_n/u_t$}.
\label{fig:Fpara}
\end{figure}

We analyze the parallel force during the first displacement and consider its average value over the length of the stroke (red dashed line in Fig. \ref{fig:signals}(b)). Figure \ref{fig:Fpara}(a) shows the measured parallel force as a function of depth for different angles $\alpha$. For a given angle, we observe that the force has a linear behavior with depth, which is consistent with a hydrostatic-type evolution of the pressure in the granular medium \cite{Zhang2014}. We observe that the slope $\gamma(\alpha)=dF_{\parallel}/dh$ increases with the angle $\alpha$ (Inset in Fig. \ref{fig:Fpara}(a)). These measurements of parallel forces are also shown as a function of the angle $\alpha$ for the different depths in Fig. \ref{fig:Fpara}(b). From these data, we estimate the parallel friction coefficient defined as $\mu_{\parallel} = F_{\parallel} / \rho_g g h A_m =\gamma(\alpha) / \rho_g g A_m$, where $g$ is the acceleration of gravity and $A_m=\pi D L$ is the mantle surface of the cylinder. Figure \ref{fig:Fpara}(c) shows the parallel friction coefficient as a function of the angle at different depths. We observe that the coefficient of friction $\mu_{\parallel}(\alpha)$ does not clearly depend on depth, since the curves collapse on the same trend, and depends solely on the orientation of the cylinder. This indicates that the drag force depends mainly on the orientation of the object.

%%%%%%%%%%%%%%%%%%%%%%%%%%%% FIGURE 3 %%%%%%%%%%%%%%%%%%%%%%%%%%%%%%%%%%%%%%%%%%%%%%%%%%%%%%%%%%%%%%%%%%%%%%%%%%%%%%%%%%%%%%
%%%%%%%%%%%%%%%%%%%%%%%%%%%%%%%%%%%%%%%%%%%%%%%%%%%%%%%%%%%%%%%%%%%%%%%%%%%%%%%%%%%%%%%%%%%%%%%%%%%%%%%%%%%%%%%%%%%%%%%%%%%%

%%%%%%%%%%%%%%%%%%%%%%%%%%%%%%%%%%%%%%%%%%%%%%%%%%%%%%%%%%%%%%%%%%%%%%%%%%%%%%%%%%%%%%%%%%%%%%%%%%%%%%%%%%%%%%%%%%%%%%%%%%%%

\subsection{Perpendicular force}\label{subsec:perp_force}
%%%%%%%%%%%%%%%%%%%%%%%%%%%% FIGURE 4 %%%%%%%%%%%%%%%%%%%%%%%%%%%%%%%%%%%%%%%%%%%%%%%%%%%%%%%%%%%%%%%%%%%%%%%%%%%%%%%%%%%%%%
\begin{figure}[t] 
\centering
\includegraphics[width=1\columnwidth]{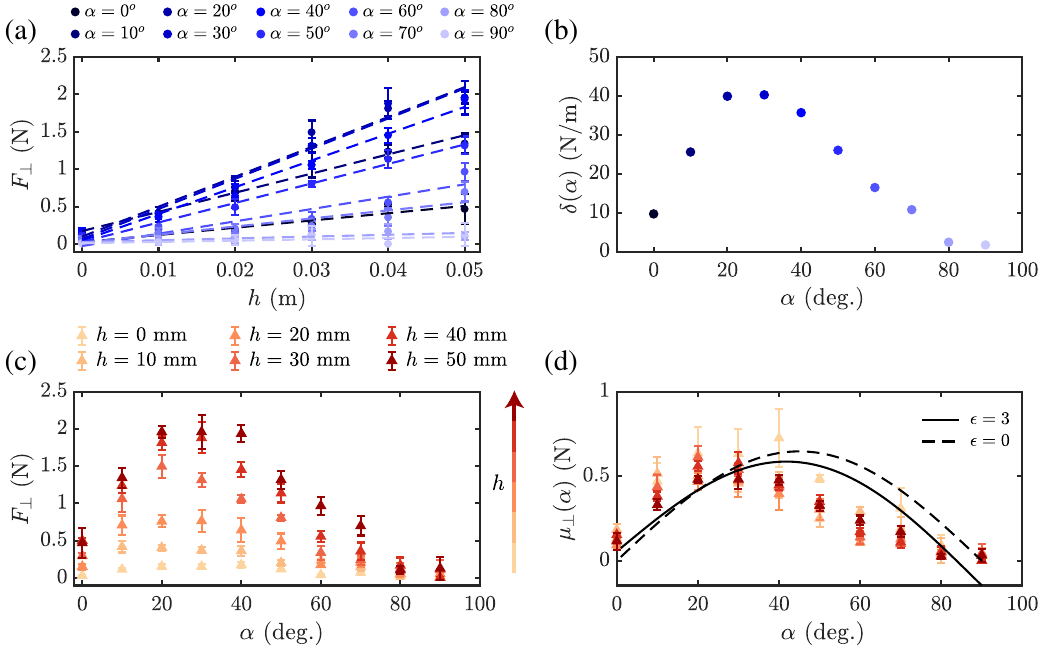}
\caption{(a) Perpendicular force as a function of depth at a constant angle. (b) Slopes of each curve as a function of angle from (a). (c) Perpendicular force as a function of angle, colormap corresponds to the respective depth. (d) Perpendicular friction coefficient as a function of angle, segmented black line corresponds to the theoretical friction coefficient without corrections (Eq. (\ref{eq8}) with $\epsilon = 0$) and solid black line for the theoretical friction coefficient with corrections (Eq. (\ref{eq8}) with $\epsilon = 3$).}
\label{fig:Fperp}
\end{figure}
%%%%%%%%%%%%%%%%%%%%%%%%%%%%%%%%%%%%%%%%%%%%%%%%%%%%%%%%%%%%%%%%%%%%%%%%%%%%%%%%%%%%%%%%%%%%%%%%%%%%%%%%%%%%%%%%%%%%%%%%%%%%
As with the parallel force, we analyze the average perpendicular force during the first displacement (red dashed line in Fig. \ref{fig:signals}(c)). Figure \ref{fig:Fperp}(a) shows the perpendicular force measurements as a function of depth. Similar to the parallel force, the perpendicular force shows a linear trend with depth; however, the slope $\delta(\alpha)=dF_{\perp}/dh$ does not increases monotonically with the angle (Fig. \ref{fig:Fperp}(b)), as was the case for the parallel force. We note that
$\delta(\alpha)$ reaches a maximum around $\alpha\approx 35^\circ$. The same data for the perpendicular forces are presented as a function of the angle $\alpha$ in Fig. \ref{fig:Fperp}(c). We estimate the friction coefficients as $\mu_{\perp}=F_{\perp}/\rho_g g A_m h=\delta(\alpha)/ \rho_g g A_m$. Figure \ref{fig:Fperp}(d) shows the perpendicular friction coefficient as a function of the orientation angle for different depths. Similar to the parallel friction coefficient, $\mu_{\perp}$ does not depend on depth, since the curves collapse on the same tendency, but moslty on the orientation of the object. The function $\mu_{\perp}(\alpha)$ is non-monotonic and reaches a maxiumum for $\alpha\approx 35^\circ$.\\
Experimentally, the force components in the parallel and perpendicular directions are measured independently and no link between both components is assumed. In the following we will develop theoretical considerations that show that these two components are intrinsically linked.

%%%%%%%%%%%%%%%%%%%%%%%%%%%%%%%%%%%%%%%%%%%%%%%%%%%%%%%%%%%%%%%%%%%%%%%%%%%%%%%%%%%%%%%%%%%%%%%%%%%%%%%%%%%%%%%%%%%%%%%%%%%%

\section{Discussion}\label{sec:model}
%%%%%%%%%%%%%%%%%%%%%%%%%%%% FIGURE 5 %%%%%%%%%%%%%%%%%%%%%%%%%%%%%%%%%%%%%%%%%%%%%%%%%%%%%%%%%%%%%%%%%%%%%%%%%%%%%%%%%%%%%%
\begin{figure}[b]
\centering
\includegraphics[width=1\columnwidth]{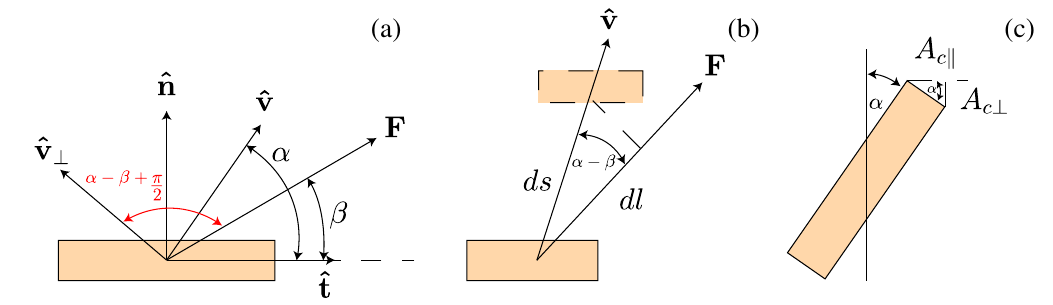}
\caption{Sketch of the analytical model showing the notation used. (a) Object experiencing a force $\mathbf{F}$ and it moves with a direction $\mathbf{\hat{v}}$. (b) Point of view using conservation of energy as it moves. (c) Correction to the theoretical model due to the effect of the cap.}
\label{fig:model}
\end{figure}
%%%%%%%%%%%%%%%%%%%%%%%%%%%%%%%%%%%%%%%%%%%%%%%%%%%%%%%%%%%%%%%%%%%%%%%%%%%%%%%%%%%%%%%%%%%%%%%%%%%%%%%%%%%%%%%%%%%%%%%%%%%%
In this section, we develop theoretical considerations about the force that applies to a cylinder when it is dragged into a granular medium. As this force depends on the orientation of the cylinder with respect to its direction of motion, we draw inspiration from the problem of friction between the snake's scales and a solid substrate, which also depends on the snake's orientation and for which theoretical elements of anisotropic friction have been introduced \cite{Hu2009}. Thus, we consider the force $\mathbf{F}$ required to overcome friction for an elementary part of the cylinder which moves along a direction given by the unit vector $\mathbf{\hat{v}}$, as illustrated in Fig. \ref{fig:model}(a). The tangential and normal directions of the cylinder are given by unit vectors $\mathbf{\hat{t}}$ and $\mathbf{\hat{n}}$ respectively. The angle between $\mathbf{\hat{t}}$ and $\mathbf{\hat{v}}$ is $\alpha$ and the angle between $\mathbf{\hat{t}}$ and $\mathbf{F}$ is $\beta$. Hu \textit{et al.} have proposed to decompose the force as $\mathbf{F}= u_t \, F_N (\mathbf{\hat{v}}\cdot \mathbf{\hat{t})} \ \mathbf{\hat{t}} + u_n \, F_N (\mathbf{\hat{v}} \cdot \mathbf{\hat{n}}) \mathbf{\hat{n}}$ where $F_N$ is the vertical force that applies on the elementary part and for which they introduced a tangential coefficient $u_t$ and a normal coefficient $u_n$ \cite{Hu2009}. In the case of a snake sliding on a solid substrate, they have shown that the tangential and normal coefficients $u_t$ and $u_n$ can be considered constant and do not depend on the orientation $\alpha$. The application of this framework to the case of an object moving in grains is similar to the granular RFT approach where effective friction coefficients are introduced and depend on the orientation of the element. The difference between $u_t$ and $u_n$ is associated with a change in the structure of the granular flow, which is not the same for tangential and normal motion. In this approach, the force component parallel to the velocity $F_{\parallel} = \mathbf{F}\cdot \mathbf{\hat{v}} = \mu_{\parallel} (\alpha) \,F_N$ writes: 
\begin{equation}\label{eq1}
    F_{\parallel}  = u_t \, F_N  (\mathbf{\hat{v}}\cdot \mathbf{\hat{t}})^2 + u_n \, F_N  (\mathbf{\hat{v}}\cdot \mathbf{\hat{n}})^2,
\end{equation}
\noindent and the component of the force that is perpendicular to the velocity $F_{\perp} = \mathbf{F} \cdot \mathbf{\hat{v}}_{\perp} =  \mu_{\perp} (\alpha) \,F_N$ writes:
\begin{equation}\label{eq2}
    F_{\perp}  = u_t \, F_N  (\mathbf{\hat{v}}\cdot \mathbf{\hat{t}}) (\mathbf{\hat{t}} \cdot \mathbf{\hat{v}}_{\perp}) + u_n \, F_N  (\mathbf{\hat{v}}\cdot \mathbf{\hat{n}}) (\mathbf{\hat{n}} \cdot \mathbf{\hat{v}}_{\perp}).
\end{equation}

\noindent We deduce that $\mu_{\parallel}= F_{\parallel}/F_N = u_t \cos^2 \alpha + u_n \sin^2 \alpha$ and $ \mu_{\perp} = F_{\perp}/F_N = (u_n - u_t) \cos \alpha \, \sin \alpha$. However, the prediction for the normal friction coefficient has been shown to differ from the prediction of Chateau \textit{et al.}, which is based on energetic considerations \cite{Chateau2013,texier2018optimal}.

Chateau \textit{et al.} assumed that the work of the external forces that are required to move the object with a displacement $d \mathbf{s}= ds \ \mathbf{\hat{v}}$ is completely dissipated by friction \cite{Chateau2013}. The energy dissipated by the effective friction force in this displacement is $\delta W = \mu_{\parallel} (\alpha) F_N \ ds$. Chateau \textit{et al.} showed that this approach has the following implications: (i) \textit{$\mu_{\parallel}(\alpha) \, F_N$ can be considered as the minimum necessary force that when applied in the direction $\alpha$, generates motion in that direction, and (ii) the direction of sliding is given by the point where the curves $\mu_{\parallel} (\alpha) \, F_N$ and $\mathbf{F} \cdot \mathbf{\hat{v}}$ are tangent} \cite{Chateau2013}. 

\noindent From the conservation of energy between the work of the applied force and the friction force, we get that $F dl = \mu_{\parallel} (\alpha) \, F_N \ ds$, where $dl=\cos(\alpha-\beta) ds$ as illustrated in Fig. \ref{fig:model}(b). The friction coefficient goes as $\mu_{\parallel} (\alpha) F_N =F \cos(\alpha-\beta)$. The work of the force $\mathbf{F}$, along a displacement $ds$ with direction $\mathbf{\hat{v}}$ is:
\begin{equation}\label{eq3}
    \mathbf{F} \cdot \mathbf{\hat{v}} ds = \mu_{\parallel}(\alpha) \, F_N ds.
\end{equation}
To calculate the derivative of Eq. \ref{eq3} with respect to the angle $\alpha$, assuming that the magnitude of the force is constant, there is only the term $\cos(\alpha-\beta)$ which depends on $\alpha$ with $d \cos(\alpha-\beta) / d\alpha = \cos(\alpha-\beta+\pi/2)$. This indicates that the derivative of the Eq. \ref{eq3} corresponds to the projection between the force and the perpendicular component of the velocity:
\begin{equation}\label{eq4}
    \frac{d(\mathbf{F}\cdot \mathbf{\hat{v}})}{d\alpha} = \frac{d\mu_{\parallel}(\alpha)}{d\alpha} \, F_N \equiv \mu_\perp \, F_N,
\end{equation}
for the case that the direction of the force is impose, we have that $d(\mathbf{F}\cdot \mathbf{\hat{v}})/d\alpha = \mathbf{F}\cdot d\mathbf{\hat{v}}/d\alpha$. Combining this assumption and Eq. \ref{eq4}, we obtain that the force component perpendicular to the velocity corresponds to the derivative of the parallel force with respect to $\alpha$. For the parallel friction coefficient, we follow the expression proposed by Hu \textit{et al.} given by Eq.  \ref{eq1}, which leads to:
\begin{equation}\label{eq5}
    \mu_{\parallel} = u_t \cos^2 \alpha + u_n \sin^2 \alpha,
\end{equation}

\noindent For the perpendicular friction coefficient, using Eqs. \ref{eq4} and \ref{eq5} with respect to $\alpha$, we get:
\begin{equation}\label{eq6}
    \mu_{\perp} = 2(u_n-u_t) \cos(\alpha) \sin(\alpha).
\end{equation}

\noindent This prediction differs by a factor two from the expression of Hu \textit{et al.} for the perpendicular friction coefficient \cite{Hu2009}. Equation (\ref{eq6}) implies that at small angles $\alpha$, the parallel friction coefficient tends to the tangential coefficient; the parallel friction coefficient increases until it reaches the value of the normal coefficient $u_n$ for $\alpha=\pi/2$. On the other hand, the perpendicular friction coefficient oscillates with the amplitude of $u_n-u_t$ and has a maximum at $\pi/4$, this indicates that the propulsion is more efficient when an object moves with this orientation. These predictions can be compared with our measurements. For each set of parallel friction coefficients measured at a given depth, we find the best fit of the data with Eq. (\ref{eq5}) by considering the two coefficients $u_t$ and $u_n$ as free parameters. This method provides estimates of the tangential and normal coefficients which are shown as a function of $h$ in Fig. \ref{fig:Fpara}(d). Both coefficients are observed to remain relatively constant and do not depend on depth, the tangential coefficient is $u_t \approx 0.39 \pm 0.08$ and the effective normal coefficient is $u_n \approx 1.04 \pm 0.05$ (red squares and blue asterisks respectively in Fig. \ref{fig:Fpara}(d)). We also present in Fig. \ref{fig:Fpara}(d) the coefficient ratio $u_n/u_t$ which first decreases with depth prior to saturate for depths $h>D$ (black diamonds in Fig. \ref{fig:Fpara}(d)). The large depth value of the ratio, $u_n/u_t\approx 1.5$, is comparable to the value observed by Hu \textit{et al.} in the case of snake skins sliding on a solid substrate \cite{Hu2009}. The prediction of the perpendicular friction coefficient based on Eq. (\ref{eq6}) and the values determined for $u_t$ and $u_n$ is represented by a solid line in Fig. \ref{fig:Fperp}(d). The theoretical prediction is in reasonable agreement with the data except for the position of the maximum which differs slightly between experiments and theory. 

\noindent To refine the model, we consider the effect of the ends of the cylinder on the force it experiences. The previous approach based on two effective coefficients $u_t$ and $u_n$ that apply along the length of the cylinder can be applied to the surface at its end. This surface has an area $A_c=\pi D^2/4$, with a normal contribution from the projected parallel area $A_{c \parallel}=A_c \cos(\alpha)$, and a tangential contribution from the projected perpendicular area $A_{c \perp}=A_c \sin(\alpha)$ (see Fig. \ref{fig:model}(c)). This induces the following correction in the expression of the parallel friction coefficient:
\begin{equation}\label{eq7}
    \mu_{\parallel} = u_t \cos^2 (\alpha) + u_n \sin^2 (\alpha) + \epsilon \frac{A_c}{A_m} (u_n \cos(\alpha)+u_t \sin(\alpha)),
\end{equation}
\noindent where the ratio $A_c/A_m \approx 5\%$ for our cylinder and $\epsilon = 3$ is a constant parameter that has been determined to best fit our measurements. The fact that the parameter $\epsilon$ is greater than one can be associated with the fact that the movement of the tip of the cylinder displaced a larger volume of grains relatively to a surface along the cylinder, in agreements with the results of DEM simulations \cite{Lehuen2020}. Then, the perpendicular friction coefficient is deduced deriving Eq. \ref{eq7} relatively to $\alpha$ in agreement with (Eq. \ref{eq4}) to get: 
\begin{equation}\label{eq8}
    \mu_{\perp} = 2(u_n-u_t) \cos(\alpha) \sin(\alpha)+\epsilon \frac{A_c}{A_m} (u_t\cos(\alpha)-u_n\sin(\alpha)).
\end{equation}
Fig. \ref{fig:Fpara}(c) shows the parallel friction coefficient without correction ($\epsilon=0$) and with correction ($\epsilon=3$). Similarly, Fig. \ref{fig:Fperp}(d) shows the perpendicular friction coefficient; the non-corrected solution (Eq. \ref{eq6}) has its maximum at $\alpha = 45^o$, and we find that the corrected solution (Eq. \ref{eq8}) shifts the curve to the left, this gives a reasonable description to the experimental data.

\noindent Undoubtedly, the complex behavior of a granular medium is currently a challenge due to the lack of universal constitutive laws, as well as the additional complexity provided by the shape of the intruder in the media. Experimental results in the literature show that granular RFT is a reasonable approximation. In contrast, micro-scale effects (e.g. shape, smoothness, and hardness) are also relevant. Askari and Kamrin show the importance of such effects as frictional plasticity \cite{askari2016intrusion}.
It is also relevant to mention the importance of the contact between particles, Sandeep and Senetakis show the influence of the roughness and Young's modulus of the particles on the coefficient of friction \cite{Sandeep2018}.
%%%%%%%%%%%%%%%%%%%%%%%%%%%%%%%%%%%%%%%%%%%%%%%%%%%%%%%%%%%%%%%%%%%%%%%%%%%%%%%%%%%%%%%%%%%%%%%%%%%%%%%%%%%%%%%%%%%%%%%%%%%%

\section{Conclusion}\label{Conclusion}

In this work, we studied the anisotropic nature of the frictional force acting on a cylinder moving into a granular medium. We measured the components of the friction force in the direction parallel and perpendicular to the velocity of the cylinder. Both components were shown to be independent of the speed of the cylinder and to increase linearly with its depth in the granular medium, allowing to introduction of friction coefficients associated with the parallel and perpendicular directions. More importantly, we observed that the parallel component of the friction is always higher than the perpendicular component, \textit{i.e.} the force is anisotropic. The parallel component increases monotonically with the angle of orientation $\alpha$ of the cylinder with its direction of velocity, while the perpendicular component reaches a maximum value for $\alpha \simeq 35^\circ$. We developed a model based on the Resistive Force Theory which introduces two friction coefficients associated with the normal and tangential directions of the intruder. This approach was completed by energetic considerations that allow to establish a relationship between the parallel and perpendicular components of the friction. The agreement between the model and experiments is better if the caps of the cylinder are taken into account in the model.

\noindent This study proves that the description of anisotropic friction forces proposed by Chateau \textit{et al.} \cite{Chateau2013}, and verified on corrugated surfaces, is also valid to account for the anisotropy of the force experienced by an object moving into a granular medium. Here, we have focused on the friction force during the first horizontal displacement of the intruder, but it would be interesting in the future to study the evolution of the force anisotropy during the following cycles of displacement or for any type of three-dimensional movement in the granular medium.

\begin{acknowledgments}
F.M. acknowledges ANID-Chile through Fondequip No. 130149. Support from LIA-MSD France-Chile (Laboratoire International Associ\'e CNRS, "Mati\`ere: Structure et Dynamique") is greatly acknowledged. We acknowledge the support from DICYT Grant 042231MH-Postdoc, Vicerrectoría de Investigación, Desarrollo e Innovación, of Universidad de Santiago de Chile. C.V. acknowledges ANID National Doctoral Program Grant No. 21201036.
\end{acknowledgments}

%%%%%%%%%%%%%%%%%%%%%%%%%%%%%%%%%%%%%%%%%%%%%%%%%%%%%%%%%%%%%%%%%%%%%%%%%%%%%%%%%%%%%%%%%%%%%%%%%%%%%%%%%%%%%%%%%%%%%%%%%%%%

% Create the reference section using BibTeX:

%\bibliography{bibliography}
%\bibliographystyle{unsrt}

\end{document}